\documentclass[12pt,a4wide,epsf]{article}
\usepackage{epsf}
\usepackage[hang,nooneline]{subfigure}
\usepackage{amsmath}

\usepackage{graphicx}
\newcommand{\insertplot}[5]{\begin{figure}
 \hfill\hbox to 0.05in{\vbox to #5in{\vfill
 \inputplot{#1}{#4}{#5}}\hfill}
 \hfill\vspace{-.1in}
 \caption{#2}\label{#3}
 \end{figure}}
 \newcommand{\inputplot}[3]{
 \special{ps: plotfile #1}
}
\newcounter{fig}   
 \newcommand{\teta}{{\eta}}
 \newcommand{\tomega}{{\omega}}
\begin{document}

\title{
Rotating Ellis Wormholes in Four Dimensions
}

\vspace{1.5truecm}
\author{
{\bf Burkhard Kleihaus$^1$},
{\bf Jutta Kunz$^1$}\\
$^1$
Institut f\"ur  Physik, Universit\"at Oldenburg\\ Postfach 2503,
D-26111 Oldenburg, Germany
}

\vspace{1.5truecm}

%\date{April 14, 2006}
\date{\today}

\maketitle
\vspace{1.0truecm}

\begin{abstract}
We present rotating wormhole solutions in General Relativity,
which are supported by a phantom scalar field. 
These solutions evolve from the static Ellis wormhole,
when the throat is set into rotation.
As the rotational velocity increases,
the throat deforms until at a maximal value of the
rotational velocity, an extremal Kerr solution is encountered.
The rotating wormholes attain a finite mass and quadrupole moment.
They exhibit ergospheres and possess bound orbits.
%While their corotating ISCO resides at the throat,
%their counterrotating ISCO resides away from the throat.
%We discuss the global charges of the wormholes including their
%quadrupole moments and analyze their ergospheres and orbits.
\end{abstract}

%\maketitle

\vfill\eject

\section{Introduction}

Lorentzian wormholes are supposed
to provide connections either between otherwise separate
universes or within a single universe.
These wormholes should not possess horizons,
like the Einstein-Rosen bridge \cite{Einstein:1935tc}
or physical singularities. 
Instead they should be smooth solutions 
of the Einstein-matter equations with a nontrivial
topology. 

The simplest such wormhole is the Ellis wormhole
\cite{Ellis:1973yv,Ellis:1979bh,Bronnikov:1973fh}.
Its throat is held open by the presence of a phantom field,
i.e., a scalar field with
a reversed sign in front of its kinetic term.
In Einstein gravity this choice of sign is necessary in order to 
provide the required violation of the energy conditions.
This was discussed in detail by Morris and Thorne 
\cite{Morris:1988cz,Visser},
whose motivation was to obtain traversable wormhole solutions
in General Relativity.
In recent years phantom fields have become ubiquitous in cosmology,
providing one of the explanations for the observed accelerated expansion
of the Universe
\cite{Lobo:2005us}.

While wormholes present a fascinating subject with their
potential impact on space travel and the possibility of time travel
\cite{Morris:1988cz,Morris:1988tu}, they may also be viewed more mundane
as potential astrophysical objects that may be searched for
observationally
\cite{Abe:2010ap,Toki:2011zu,Takahashi:2013jqa}.
In this respect, wormholes have for instance
been considered as gravitational
lenses, as first addressed in \cite{Cramer:1994qj,Perlick:2003vg},
whose Einstein rings 
\cite{Tsukamoto:2012xs}
or shadows 
\cite{Bambi:2013nla,Nedkova:2013msa}
have been studied.
But also neutron star--wormhole systems and their possible
astrophysical signatures have been addressed
\cite{Dzhunushaliev:2011xx,Dzhunushaliev:2012ke,Dzhunushaliev:2013lna,Dzhunushaliev:2014mza}.

Astrophysical objects generically carry spin. 
However,
all of the smooth wormhole solutions 
of the Einstein-matter equations
obtained so far (in four
space-time dimensions), are either static
or slowly rotating perturbative solutions 
\cite{Kashargin:2007mm,Kashargin:2008pk}\footnote{Note,
that the Teo wormhole \cite{Teo:1998dp} 
is not a solution of the Einstein-matter equations.}.
Thus we here present for the first time
a family of globally regular stationary 
wormhole solutions. These represent
the rotating generalizations of the Ellis wormholes.
We determine their domain of existence
and analyze their physical properties.
We note, that the family of rotating wormholes
ends in an extremal Kerr black hole,
%Comparing with other known types of wormholes,
and conclude, that it is a generic feature
of families of wormholes in Einstein gravity to end in 
extremal black hole solutions.

\section{The model}

%After discussing the action, the Ans\"atze and the equations of motion,
%we present the wormhole solutions and analyze their properties,
%including their embeddings and violations of the energy conditions.

We consider Einstein gravity coupled to a phantom field $\Phi$
with action
\begin{equation}
S=\int \left[ \frac{c^4}{16\pi G}R 
%+  L_{\rm ph} \right] 
 + \frac{\hbar^2}{2 m_0} \partial_\mu \Phi \partial^\mu \Phi \right]
\sqrt{-g} d^4x
\ , \label{action}  \end{equation}
where $R$ denotes the curvature scalar,
$G$ is Newton's constant, $g$ is the determinant of the metric,
and $m_0$ represents a mass scale.

To construct stationary rotating wormhole solutions of this action
we employ the line element
\begin{equation}
ds^2 = -e^{f} c^2 dt^2 +p^2 e^{-f} 
\left( e^{\nu} \left[d\teta^2 +h d\theta^2\right]
                    + h \sin^2\theta \left(d\varphi -\tomega dt\right)^2\right) \ ,
\label{lineel}
\end{equation}
where  $f$, $p$, $\nu$ and $\tomega$ are functions only of
$\teta$ and $\theta$, and $h=\teta^2 + \teta_0^2$ 
is an auxiliary function. 
We note, that the coordinate $\teta$ takes positive and negative
values, i.e.~$-\infty< \teta < \infty$. 
The limits $\teta\to \pm\infty$
correspond to two distinct asymptotically flat regions.
The phantom field $\Phi$ depends only on the coordinates
$\teta$ and $\theta$, as well.

Substitution of these Ans\"atze into the Einstein equations leads to
a system of non-linear partial differential equations (PDEs).
%to be solved numerically.
%
The PDE for the function $p$ decouples and has a simple form
\begin{equation}
p_{,\teta,\teta} +\frac{3\teta}{h}p_{,\teta} + 
\frac{2 \cos\theta}{h\sin\theta}p_{,\theta}+
\frac{1}{h}p_{,\theta,\theta} = 0 \ .
\label{pde_p}
\end{equation}
A solution which satisfies the boundary conditions 
$p(\teta \to \infty) =p(\teta \to -\infty) = 1$ and 
$p_{,\theta}(\theta =0) =p_{,\theta}(\theta =\pi) = 0$ 
is given by $p=1$.

For the phantom field we then find with $p=1$ the PDE
\begin{equation}
\Phi_{,\teta,\teta} +\frac{2\teta}{h}\Phi_{,\teta} + 
\frac{\cos\theta}{h\sin\theta}\Phi_{,\theta}+
\frac{1}{h}\Phi_{,\theta,\theta} = 0 \ .
\label{pde_phi}
\end{equation}
An explicit solution is $\Phi_{,\teta}=D/h$,
where the constant $D$ represents the scalar charge of the wormhole solution.

The remaining equations yield a set of 3 PDEs for the functions 
$f$, $\nu$ and $\tomega$ together with 
an equation for the scalar charge and a constraint equation.
%Introducing the dimensionless coordinate $\eta = \teta/r_0$
%and the dimensionless function $\omega = \tomega r_0/c$,
We solve these PDEs numerically, 
while monitoring the fulfillment of the constraint
and the constancy of the scalar charge.

We here focus on symmetric wormholes, where
$f$ and $\nu$ are even functions of $\teta$.
Their throat is located at the hypersurface $\teta=0$, 
which represents a minimal surface.
The equatorial radius $R$ of the throat is given by
\begin{equation}
R = \left. \sqrt{g_{\varphi\varphi}}\right|_{\teta=0,\theta=\pi/2}
=e^{-f_0/2} \teta_0 \ ,
\label{Req}
\end{equation}
with $f_0=f(\teta=0,\theta=\pi/2)$,
while the polar radius $R_p$ and the areal radius $R_A$ are obtained via
\begin{equation}
R_p  =\frac{\teta_0}{\pi} 
\int_0^\pi {\left. e^{(\nu-f)/2}\right|_{\teta=0}  d\theta} \ , \ \ \
R_A^2=\frac{\teta_0^2}{2}   
\int_0^\pi {\left. e^{\nu/2-f}\right|_{\teta=0} \sin \theta  d\theta} \ .
\end{equation}
Denoting the angular velocity of the throat by
$\Omega=\tomega_0$, %=\tomega(\teta=0)$,
the dimensionless rotational velocity
of the throat is given by
\begin{equation}
%\frac{\Omega}{\Omega_0}=\omega_0 e^{-f_0/2} \ , \ \ \
v_e= \frac{R\Omega}{c} \ . % =\omega_0 e^{-f_0/2} \ , \ \ \
%{\rm with}\ \ \Omega_0=\frac{c}{R} \ 
\end{equation}
%with $\Omega_0={c}/{R}$ and $\omega_0=\omega(\teta=0)$.

The mass $M$ and angular momentum $J$ of the wormhole
solutions are obtained from the asymptotic
form of the metric tensor components
\begin{equation}
g_{tt} \underset{\teta \to \infty} 
\longrightarrow - c^2\left(1-\frac{2GM}{c^2\teta}\right)  \ , \ \ \
%& \Longleftrightarrow & 
%f \to -\frac{2GM}{c^2 r_0 \eta} = -\frac{2\mu}{\eta} 
%\label{mass}\\
g_{t\varphi} \underset{\teta \to \infty} 
\longrightarrow -\frac{2G J}{c^2}\frac{\sin^2\theta}{\teta}
%& \Longleftrightarrow & 
%\omega \to -\frac{2G J}{c^3 r_0^2}\frac{1}{\eta^3}=-\frac{j}{\eta^3} 
\label{mass_angmom}
\end{equation}

In order to establish a relation between the  mass and 
the angular momentum, i.e., a Smarr-type relation,
we consider the integrals
\begin{eqnarray}
I_M  & = & \frac{c^4}{4\pi G}\int_\Sigma R_{\mu\nu} n^\mu \xi^\nu dV
=\frac{1}{8\pi G}\int{\left.\left[ c^2 h f_{,\teta} \sin\theta
                            -h^2 \tomega\tomega_{,\teta} e^{-2 f}\sin^3\theta 
                        \right]\right|_0^\infty d\theta d\varphi} \ ,
\label{im}\\
I_J  & = & -\frac{c^4}{8\pi G}\int_\Sigma R_{\mu\nu} n^\mu \psi^\nu dV
=\frac{1}{16\pi G}\int{\left. \left[c^3 h^2 \tomega_{,\teta} e^{-2 f}\sin^3\theta 
                        \right]\right|_0^\infty d\theta d\varphi} \ .
\label{ij}
\end{eqnarray}
Here $\Sigma$ is a spatial hypersurface extending from the throat $\teta=0$
to the asymptotically flat region $\teta \to \infty$,
$dV$ denotes its natural volume element and
$n^\mu$ is normal on $\Sigma$.
$\xi^\nu$ is the time-like Killing vector field normalized to one in the
asymptotically flat region $\teta \to \infty$, and $\psi^\nu$ is the
space-like Killing vector field. 
In evaluating the expressions Eqs.~\ref{im}-\ref{ij},
we have taken into account
that the solutions are regular on the symmetry axis and possess reflection
symmetry with respect to the equatorial plane.
On the other hand the integrals have to vanish as a consequence of the
Einstein equations. This yields for Eq.~(\ref{ij})
\begin{equation}
\int{c^3 \left[h^2 \tomega_{,\teta} e^{-2 f}\right]_{\teta=0}
\sin^3\theta  d\theta d\varphi}
= - 16 \pi c G J \ ,
\end{equation}
while substitution of this result in
Eq.~(\ref{im}) leads to the desired mass relation
\begin{equation}
M c^2 = 2 \tomega_0 J \ \ \ \Longleftrightarrow \ \ \
\frac{M}{M_0} = \frac{\Omega}{\Omega_0}\frac{J}{J_0} \ ,
\label{smarr}
\end{equation}
%\frac{M}{M_0}= \mu e^{f_0/2} \ , \ \ \
%\frac{J}{J_0}= j e^{f_0} \ , \ \ \
%M_0= \frac{Rc^2}{G} \ , \ \ \ J_0= \frac{R^2c^3}{2G} \ .
where $M_0= {Rc^2}/{G}$, $\Omega_0={c}/{R}$ and $J_0= {R^2c^3}/{2G}$.

%Finally, for the phantom field we define dimensionless scalar charge and
%energy density in the equatorial plane of the throat,
%\begin{equation}
%\frac{q^2}{q_0^2}= D^2 e^{f_0}\ , \ \ \
%\frac{\epsilon}{\epsilon_0} = -\frac{D^2}{2}e^{-\nu_0}\ , \ \ \
%{\rm with} \ \ q_0^2 = \frac{c^4 m_0 R^2}{4\pi G \hbar^2}\ , \ \ \
%\epsilon_0= \frac{c^4}{4\pi R^2 G} \ ,
%\end{equation}
%where $D^2$ denotes the rhs of Eq.~(\ref{q2}).

To better understand the wormhole solutions 
we now turn to the discussion of the second asymptotic limit,  
$\teta \to -\infty$.
As already noted for the perturbative solutions
\cite{Kashargin:2007mm,Kashargin:2008pk},
the function
$\tomega$ cannot tend to zero in both asymptotically flat regions,
$\teta \to \infty$ and $\teta \to -\infty$.
To see this in the general case,
we express the PDE for $\tomega$
in the form of a conservation law
\begin{equation}
 \left(h^2 e^{-2f} \sin^3\theta \tomega_{,\teta}\right)_{,\teta}
+\left(h   e^{-2f} \sin^3\theta \tomega_{,\theta}\right)_{,\theta} = 0
\label{omeq2}
\end{equation}
and integrate over the full range of $\teta$ and $\theta$.
%$-\infty < \teta  < \infty$, $0 < \theta < \pi $. 
Since
$\tomega_{,\theta}(0)=0=\tomega_{,\theta}(\pi)$ we are left with
\begin{equation}
\int_0^{\pi}\left[h^2 e^{-2f} 
\sin^3\theta \tomega_{,\teta}\right]_{\teta\to -\infty}
d \theta = 
\int_0^{\pi}\left[h^2 e^{-2f} 
\sin^3\theta \tomega_{,\teta}\right]_{\teta\to \infty}
d \theta %\ .
\propto J \ .
\label{omeqi2}
\end{equation}
Hence, both limits are
regular and non-vanishing for rotating wormholes.
Next we multiply Eq.~(\ref{omeq2}) by $\tomega$ and integrate again. 
This yields
\begin{equation}
\int_0^{\pi}\left[h^2 e^{-2f} \sin^3\theta 
\tomega\tomega_{,\teta}\right]^{\teta\to \infty}_{\teta\to -\infty}
d \theta 
=
\int\left[h^2 e^{-2f} \sin^3\theta 
\left(\tomega_{,\teta}^2+\frac{1}{h}
\tomega_{,\theta}^2\right)\right]d\teta d\theta > 0 \ .
\label{omeqi3}
\end{equation}
Since the term $h^2 e^{-2f} \tomega_{,\teta}$ is bounded (in magnitude)
from above in both limits, 
the function $\tomega$ has to be non-zero in (at least) 
one of the asymptotic regions.
We note, that one can always arrange $\tomega(\infty)=0$ 
by adding a suitable constant.
In this case $\tomega(-\infty)= 2 \tomega_0$.
Still, the region $\teta \to -\infty$
is asymptotically flat, as one can see by means of
a simple coordinate transformation.
In that case, $\tomega(\infty)\to - 2 \tomega_0$.

\section{Numerical results}

\begin{figure}[t!]
\begin{center}
%\hspace{-2.0cm}
\mbox{\hspace{0.2cm}
\subfigure[][]{\hspace{-1.8cm}
\includegraphics[height=.28\textheight, angle =0]{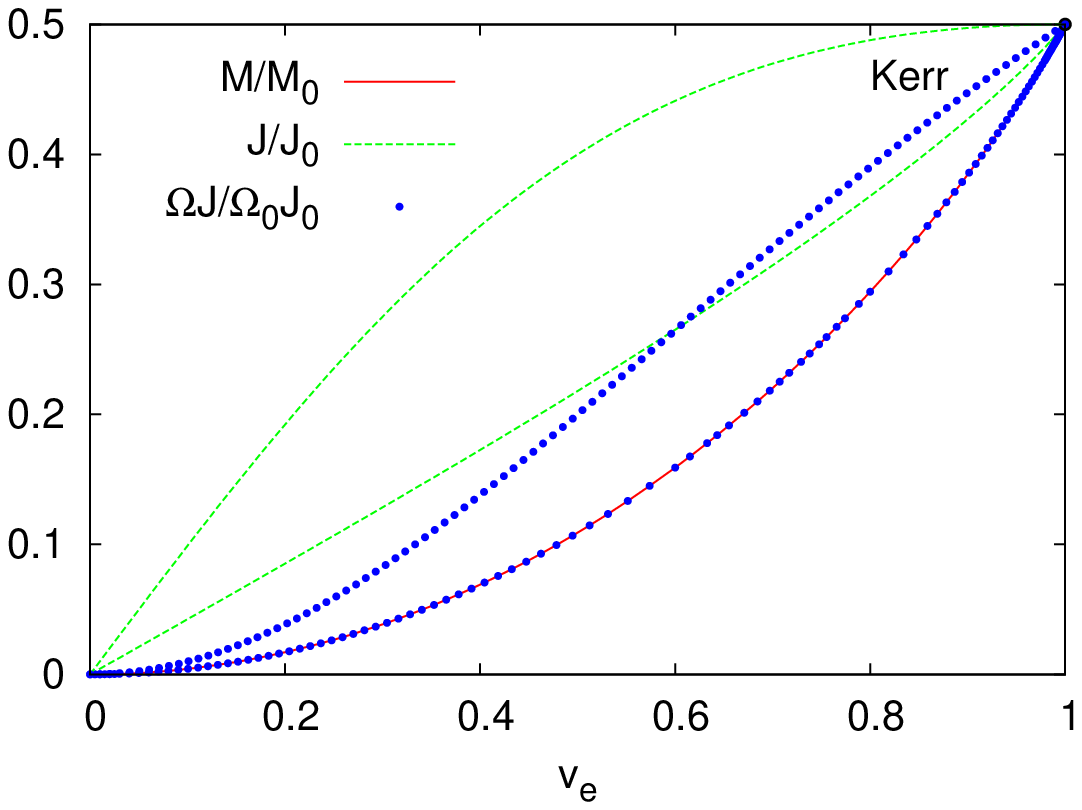}
\label{Fig1a}
}
\subfigure[][]{\hspace{-0.8cm}
\includegraphics[height=.28\textheight, angle =0]{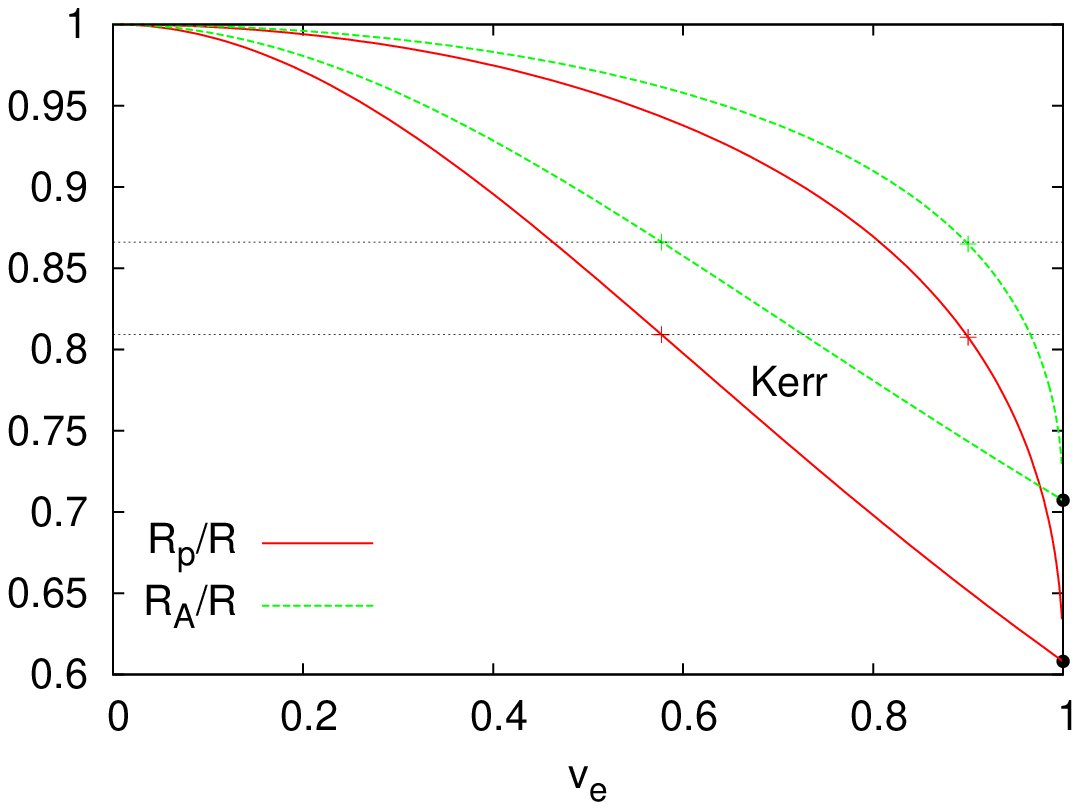}
\label{Fig1b}
}
}
\end{center}
\vspace{-0.5cm}
\caption{
(a) The dimensionless mass $M/M_0$ and angular momentum $J/J_0$ 
%and a confirmation of the Smarr-type relation 
versus the dimensionless rotational velocity $v_e$
for symmetric wormhole solutions.
Validity of the Smarr-type relation Eq.~(\ref{smarr}) is also 
demonstrated for the wormhole solutions.
(b) The polar radius $R_p/R$ and 
areal radius $R_A/R$ for fixed equatorial radius $R$
versus $v_e$.
The crosses mark the onset of a negative Gaussian curvature at the poles.
The black dots at $v_e=1$ in (a) and (b) correspond to the values
of the extremal Kerr black hole.
For comparison the respective curves for the full set of Kerr solutions
are also shown (note that $M/M_0=1/2$ in (a)).
\label{Fig1}
}
\end{figure}

We exhibit in Fig.~\ref{Fig1a} the dimensionless
mass and angular momentum
of the family of symmetric rotating wormholes, evolving from the static
Ellis wormhole. % ($M=0$).
As the dimensionless rotational velocity $v_e$ reaches its maximal value,
$v_e=1$,
the hypersurface $\teta=0$ changes its character
and a degenerate horizon forms.
The limiting configuration represents an extremal Kerr black hole.

The scalar charge on the other hand is a monotonically
decreasing function of $v_e$, which vanishes in the limit
$v_e=1$, as it should for a Kerr black hole.
Indeed, the phantom field vanishes identically in the limit.
Accordingly, the Smarr-type relation (\ref{smarr})
holds for Kerr black holes in the extremal
limit, when $\tomega_0$ denotes the horizon angular velocity.

%We note, that it seems to be a generic feature,
%that the families of generalized 
%Ellis wormholes end in an extremal black hole.
%In particular, charged static Ellis wormholes in 4 dimensions
%end in an extremal Reissner-Nordstr\"om black hole
%\cite{},
%while rotating Ellis wormholes in 5 dimensions
%end in an extremal Myers-Perry black hole
%\cite{}.

When the spherically symmetric Ellis wormhole is set into rotation,
the throat deforms. In particular, as seen in Fig.~\ref{Fig1b},
the ratio of the polar radius to the equatorial radius $R_p/R$ decreases
monotonically and assumes the value of the ratio
of the corresponding extremal Kerr horizon radii in the limit $v_e \to 1$.
The same holds for the areal radius $R_A$.
Beyond a certain value of $v_e$ the Gaussian curvature
turns negative at the poles of the throat, as indicated
by the crosses in the figure. 
Interestingly, this occurs for the same value of the ratio $R_p/R$
for the wormholes as for the Kerr black holes, and likewise for 
the ratio $R_A/R$.

Following \cite{Kashargin:2008pk} we consider
the violation of the null energy condition (NEC) 
by evaluating the quantity $\Xi = R_{\mu\nu} k^\mu k^\nu$
with null vector 
$k^\mu = \left(e^{-f/2}, e^{f/2-\nu/2}, 0, \omega e^{-f/2}\right)$.
Since $\Xi$ is non-positive the NEC is violated everywhere.
Introducing a global scale invariant measure for the NEC violation,
$ Y = \frac{1}{R}\int{\Xi \sqrt{-g} d\eta d\theta d\varphi} $,
we observe a monotonic decrease of
$|Y|$ with increasing rotation velocity $v_e$.

\begin{figure}[t!]
\begin{center}
%\vspace{0.5cm}
\mbox{\hspace{0.2cm}
\subfigure[][]{\hspace{-1.0cm}
\includegraphics[height=.28\textheight, angle =0]{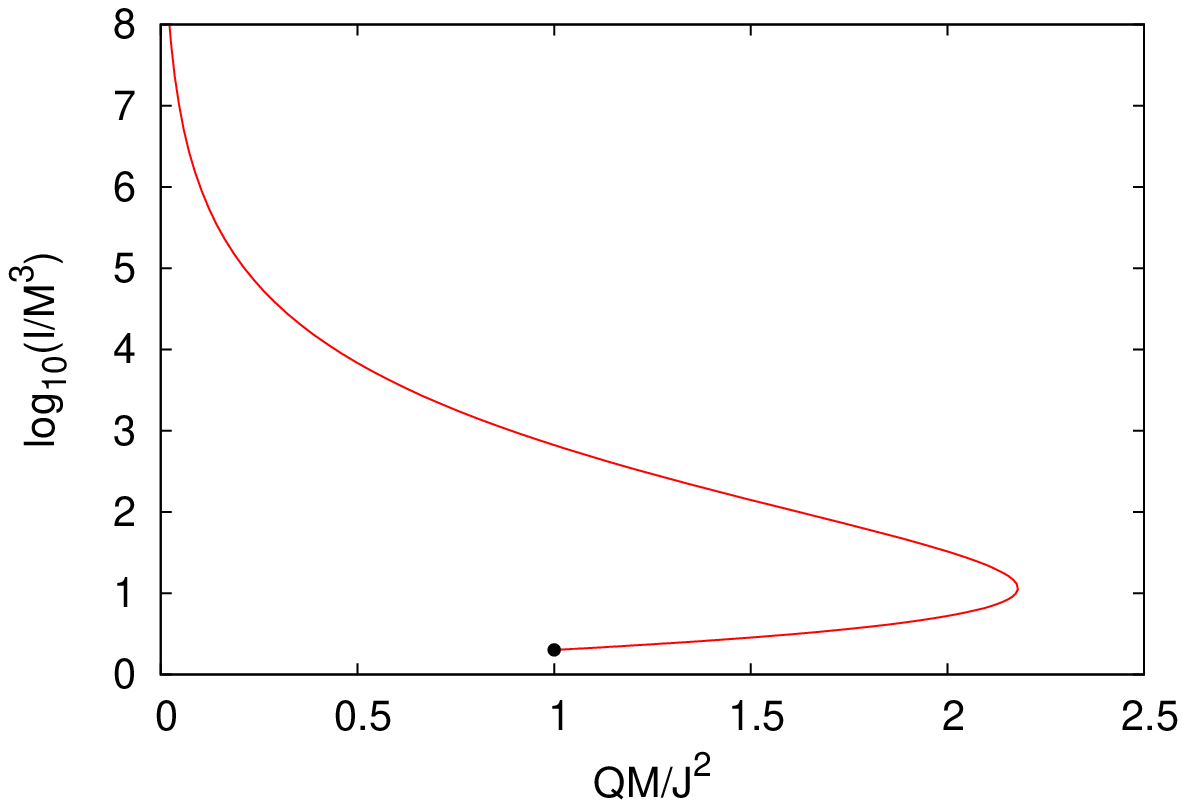}
\label{Fig2a}
}
\subfigure[][]{\hspace{-0.5cm}
\includegraphics[height=.28\textheight, angle =0]{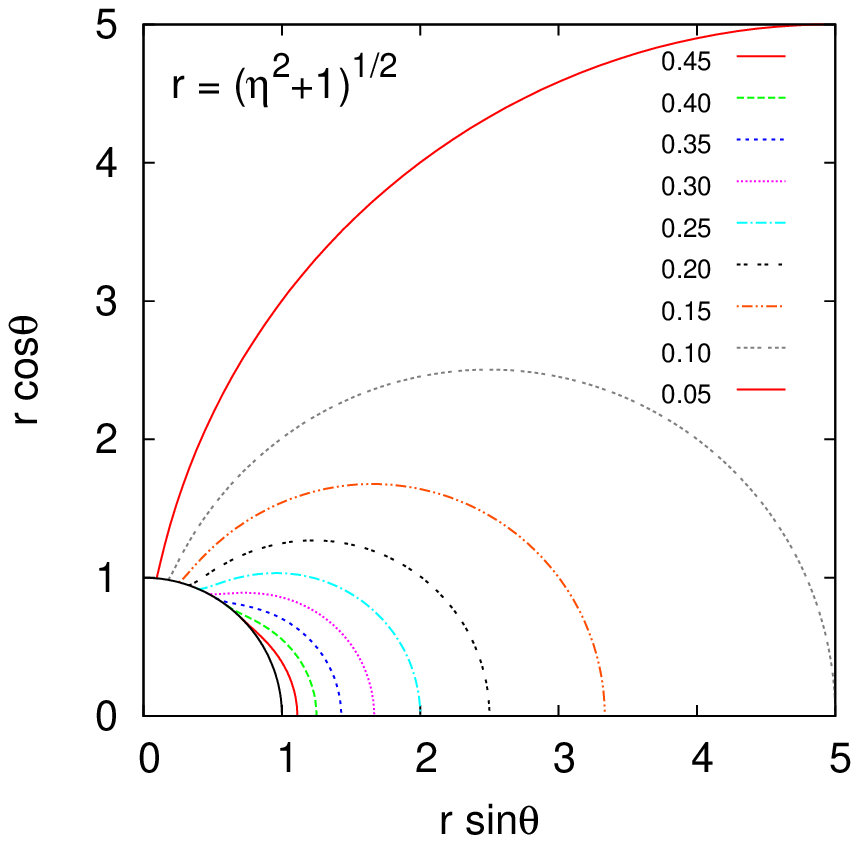}
\label{Fig2b}
}
}
\end{center}
\vspace{-0.5cm}
\caption{
\label{Fig2}
(a) 
%The dimensionless quadrupole moment $QM/J^2$ 
%versus the dimensionless rotational velocity $v_e$
%for symmetric wormhole solutions.
The dimensionless moment of inertia $I/M^3$
versus the dimensionless quadrupole moment $QM/J^2$.
The black dot at $\nu_e=1$ corresponds to the value
of the extremal Kerr black hole.
%The Kerr values correspond to the short vertical line.
(b) Quadrant of the ergoregion and the throat 
in coordinates $r$ and $\theta$
%$(r\cos\theta,r\sin\theta)$  
for fixed $\varphi$, $\teta_0=1$ and several values of $\tomega_0$.
}
\end{figure}

To extract the quadrupole moment $Q$ of the rotating wormhole solutions
we follow Geroch and Hansen \cite{Geroch:1970cd,Hansen:1974zz}
and extract it from the asymptotic expansion
in appropriate coordinates (see e.g.~\cite{Kleihaus:2014lba}).
%\begin{equation}
%Q 
%=  -M_2 +\frac{4}{3}\left[\frac{1}{4}+\frac{D_1}{M^2}
%+\frac{q^2}{16M^2}\right] M^3 .
%\label{Q}
%\end{equation}
We recall, that the dimensionless quadrupole moment $QM/J^2$ of
the Kerr solutions is constant, $QM/J^2=1$. 
The dimensionless quadrupole moment $QM/J^2$ of the 
family of rotation wormhole solutions is shown in Fig.~\ref{Fig2a}. 
It is smaller than the Kerr value for slow rotation while
it exceeds the Kerr value for fast rotation.
In the limit $v_e \to 1$, however,
the Kerr value is attained.
The dimensionless moment of inertia of the Kerr solutions
is bounded and given by ${I/M^3}_{\rm Kerr}=2(1+\sqrt{1-J^2/M^4})$.
%decreases from 4 to 2 as the rotational velocity increases.
For rotating wormholes, however, ${I/M^3}$ rises without bound 
in the limit of slow rotation, 
since the static Ellis wormhole has $M=0$.

Rotating wormholes possess an ergoregion, 
defined by the condition $g_{tt}> 0$,
%From the line element Eq.~(\ref{lineel}) we find
\begin{equation}
g_{tt}= -e^{f} c^2 +e^{-f} h \sin^2\theta \tomega^2 > 0 \ .
\label{ergoreg}
\end{equation}
While the rotation axis %(i.e.~$\sin\theta=0$)
is excluded from the ergoregion, 
its boundary in the equatorial plane is closely
related to the angular velocity $\omega_0$ of the throat.
The numerical calculations yield with high accuracy the relation
$\arctan(\teta_{\rm ergo}/\teta_0)= \arccos(2 \omega_0)$.
Thus an ergoregion exists in the asymptotically flat region,
where $\tomega \to 0$, only for sufficiently fast rotation,
while there always exists an ergoregion
in the asymptotically flat region, where $\tomega$ 
tends to a finite value. 
The ergoregion is exhibited in Fig.~\ref{Fig2b}.

\begin{figure}[t!]
\begin{center}
\includegraphics[height=.28\textheight, angle =0]{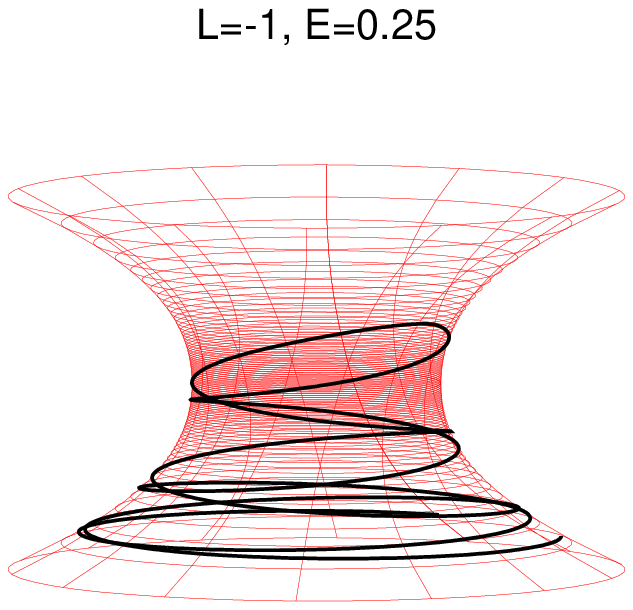}
\end{center}
\vspace{-0.5cm}
\caption{
\label{Fig3}
Embedding diagram of
the bound orbit of a test particle in the equatorial plane of a rotating
wormhole ($v_e=0.966$).
}
\end{figure}

In contrast to the static Ellis wormhole, rotating wormholes
possess stable bound orbits. Indeed, there are two types of bound orbits,
those that always remain within a single universe, and those that
oscillate between the two universes. An example of such a
two-world bound orbit is seen in Fig.~\ref{Fig3}.
For corotating orbits the innermost stable circular orbit (ISCO)
resides at the throat. For counterrotating orbits, however,
the radius of the ISCO depends on the rotation velocity,
with the ratio $R_{\rm ISCO}/R$ increasing almost by a factor of two
from the static limit to the extremal Kerr black hole.
At the same time the ratio of the orbital period $T_{\rm ISCO}/T$,
where $T$ denotes the orbital period of the throat (or the horizon of the
extremal black hole),
increases roughly by a factor of three.

\section{Conclusions}

While wormholes with phantom fields have been studied in numerous
respects (see e.g.~\cite{Visser}), we have here provided the first
globally regular rotating wormhole solutions and analyzed their
physical properties, such as their global charges including
their quadrupole moments and moments of inertia, as well as 
the properties of their throats. 
In particular, we obtained a Smarr-type relation for these wormholes.

With increasing rotation velocity the violation of the NEC decreases.
The family of rotating wormholes then ends in an extremal Kerr black hole.
We note, that it seems to be a generic feature,
that the families of generalized 
Ellis wormholes end in an extremal black hole.
In particular, charged static Ellis wormholes in 4 dimensions
end in an extremal Reissner-Nordstr\"om black hole
\cite{Hauser:2013jea},
while rotating Ellis wormholes in 5 dimensions
end in an extremal Myers-Perry black hole
\cite{Dzhunushaliev:2013jja}.

While the solutions presented are symmetric, in general
rotating wormholes need not be symmetric. By choosing non-symmetric
boundary conditions for the metric function $f$, families of
rotating wormholes emerge, which have different masses associated
to their two asymptotically flat regions,
as manifest already in the static case.
As the asymmetry increases, the wormhole properties change.
%The Smarr-type formula receives another piece, while...
But no matter how asymmetric the solutions are, these families always
appear to end in an extremal Kerr black hole.

Concerning the stability of the wormhole solutions,
it is known that the static Ellis wormholes are unstable
\cite{Shinkai:2002gv,Gonzalez:2008wd,Gonzalez:2008xk}.
However, it has been argued, that rotating wormholes 
might be stable \cite{Matos:2005uh}.
Indeed, a stability analysis of five-dimensional
rotating wormholes has shown, that the unstable mode
of the static solutions disappears, when the rotation
is sufficiently fast 
\cite{Dzhunushaliev:2013jja}.
Whether this will happen as well in four dimensions
still needs to be shown.
On the other hand, stability might also be achieved
by allowing higher curvature terms in the action
\cite{Kanti:2011jz,Kanti:2011yv}.
%See also $f(R)$.

\subsection*{Acknowledgment}

We gratefully acknowledge support 
by the DFG Research Training Group 1620 ``Models of Gravity''.

%\end{document}

\end{document}